\documentclass[aps,pra,twocolumn]{revtex4}

\usepackage{epsfig}
\usepackage{psfig}
\usepackage{bm}

\def\ie{{\it i.e.}}
\def\eg{{\it e.g.}}
\def\Erot{{E_{\rm rot}}}
\def\E2rot{{E^\prime_{\rm rot}}}
\def\ErotMM{{E^{\prime\prime}_{\rm rot}}}
\def\energy{{E}}

\begin{document}

\title{Single vortex states in a confined Bose-Einstein condensate}
\author{S. Komineas, N.R. Cooper}
\affiliation{
Theory of Condensed Matter Group,  Cavendish Laboratory,
Madingley Road, Cambridge CB3 0HE, United Kingdom}
\author{N. Papanicolaou}
\affiliation{
Department of Physics, University of Crete, and Research Center of Crete,
Heraklion, Greece}

\date{\today}

\begin{abstract}
It has been demonstrated experimentally
that non-axially symmetric vortices precess
around the centre of a Bose-Einstein condensate.
Two types of single vortex states have been observed, usually referred to
as the S-vortex and the U-vortex.
We study theoretically the single vortex excitations in spherical
and elongated condensates as a function of the interaction strength.
We solve numerically the Gross-Pitaevskii equation and
calculate the angular momentum as a function of precession frequency.
The existence of two types of vortices means that
we have two different precession frequencies for each angular
momentum value.
As the interaction strength increases the vortex lines
bend and the precession frequencies shift to lower values.
We establish that for given angular momentum the S-vortex has higher energy than the U-vortex in a rotating
elongated condensate. We show that the S-vortex is related to the solitonic vortex
which is a nonlinear excitation in the nonrotating system.
For small interaction strengths the S-vortex is related to the dark soliton.
In the dilute limit a lowest Landau level calculation provides an
analytic description of these vortex modes in terms of the harmonic
oscillator states.
\end{abstract}

\maketitle

\section{Introduction}   \label{sec:intro}

Quantised vortices in superfluids have been the subject of
an important research area
since their theoretical prediction in the context of liquid helium
\cite{donnelly}.
In the last years there has been intense experimental and theoretical study
on the creation and dynamics of
single vortices as well as vortex lattices in 
ultracold atomic Bose-Einstein condensates (BECs)
\cite{rosenbusch,abo-shaeer}.
These systems offer the possibility for direct observation of vortex lines
and of their dymamics.
They are thus excellent systems for a comprehensive theoretical
study of vortex lines in three dimensions.

In an experiment performed in a spherical trap \cite{anderson}
single vortices were produced by a technique which induces
a $2\pi$ phase winding using a rotating laser beam.
The vortices were located off the trap center,
and they were seen to precess robustly around it at an almost
constant angular velocity.
Surprisingly the precession frequency was found to be
almost independent of the displacement of the vortex from the trap center.

In another important experiment it was established that
the vortex lines in the finite trap geometry
can have nontrivial shapes \cite{rosenbusch}.
It was observed that a single vortex line may have an S shape or
a U shape.
The length and precise shape of the vortex can vary and this
is directly related to the angular momentum it carries.
Theoretical work in high density very elongated condensates
similar to those of Ref.~\cite{rosenbusch}
showed that the U-shaped vortex can be
the ground state of the rotating system \cite{garcia-ripoll,modugno}.
Some differences between the S and the U-vortex
were reported in \cite{aftalion}.

In this paper, we establish that an axisymmetric BEC generically supports both S and U-shaped vortices.
We study these single vortices as precessing
stationary states within the nonlinear
Gross-Pitaevskii theory for a range of interaction strengths.
Their frequency of precession around the trap
center depends critically on the features of the vortex line.
Precession frequencies for U-vortices depend only weakly on their
angular momentum and fall in a relatively
narrow frequency range which depends on the interaction strength.
S-vortices have higher energy and their
precession frequencies depend strongly on their angular
momentum. The nonrotating member of this family is
a non-axially symmetric vortex solution of the nonlinear theory.
This has been called a {\it solitonic vortex}; it breaks
spontaneously the symmetry in an axially symmetric potential
and pierces through the trap in a direction perpendicular to
the symmetry axis \cite{brand,komineas3}.

In the dilute limit the low-energy states
can be described within a lowest Landau level (LLL) ansatz \cite{butts}.
Motivated by the fact that experiments are usually performed in elongated
condensates we generalize the LLL ansatz to include longitudinal modes.
Both types of single vortex states can then be described within
this formalism and some simple analytic expressions capture the
main features of the vortex modes.
Low density condensates close to the LLL regime 
have been created recently by fast rotation where a vortex lattice
was formed \cite{schweikhard}.

The outline of the paper is as follows.
In Sec.~II we formulate the problem and
find numerically two types of
single vortex solutions of the Gross-Pitaevskii equation.
We apply our method for a spherical trap as in
\cite{anderson} as well as for elongated condensates.
In Sec.~III we formulate the problem in the LLL augmented to include
the longitudinal modes and we study the
S and U-vortices within this formalism. Finally, Sec.~IV contains
a summary and our concluding remarks.

\section{Precessing vortices}

Motivated by the experiment \cite{anderson} where off-center vortices
precess at an almost constant angular velocity around the
center of the condensate, we look here for
vortices which are stationary states in a frame rotating at a constant 
frequency.
In this paper we suppose that the energy of a BEC is given by the
Gross-Pitaevskii theory.
We consider an axially symmetric harmonic potential
with frequencies $\omega_\perp$ in the transverse direction and
$\omega_\|$ in the longitudinal direction and
implement rationalised units by the substitutions
\begin{equation}  \label{eq:units}
 x \to a_\perp\,x, \quad y \to a_\perp\,y, \quad z \to a_\|\,z, \quad
 \Psi \to \frac{N^{1/2}}{a_\perp  a_\|^{1/2}}\,\Psi,
\end{equation}
where $\Psi$ is the condensate wave function,
$N$ is the number of atoms in the condensate,
and $a_{\perp,\|}=\sqrt{\hbar/m\omega_{\perp,\|}}$ are
the oscillator lengths in the transverse and longitudinal
directions. The wave function normalisation condition reads
\begin{equation}  \label{eq:normalization}
\int{|\Psi|^2\, dV} = 1.
\end{equation}
The energy functional per particle in units of $\hbar\omega_\perp$
has the form
\begin{eqnarray}  \label{eq:energy}
 \energy  & = &
 \frac{1}{2}\, \int  \left[
 \left| \bm{\nabla}_\perp \Psi\right|^2
+ \rho^2 |\Psi|^2  \right.  \nonumber \\
& + & \beta\, \left[  \left|\frac{\partial \Psi}{\partial z}\right|^2
 +  z^2 |\Psi|^2 \right]
 \left. + 4\pi \frac{N a_s}{a_\|}\, |\Psi|^4
\right]
dV ,
\end{eqnarray}
where $\rho,z,$ and $\phi$ are cylindrical coordinates,
$\bm{\nabla}_\perp$ is the gradient on the transverse plane,
$a_s$ is the scattering length,
and the parameter
\begin{equation}  \label{eq:beta}
 \beta \equiv  \frac{\omega_\|}{\omega_\perp}
\end{equation}
is the ratio of the two frequencies.
The value $\beta=1$ corresponds to a spherical trap.
The quartic term gives the strength of the interactions
and we introduce for later convenience the parameter
\begin{equation}  \label{eq:delta}
 \delta \equiv  \frac{1}{2 \sqrt{2\pi}}\,\frac{N a_s}{a_\|}
\end{equation}
which is a dimensionless measure of the interaction strength relative to the transverse trapping energy $\hbar\omega_\perp$.

Stationary states which precess at a constant angular frequency $\omega$
are extrema of the energy in the rotating frame:
\begin{equation}  \label{eq:E-omegaL}
  \Erot = \energy - \omega \ell,
\end{equation}
where $\ell$ is the angular momentum per particle along the symmetry axis,
in units of $\hbar$:
\begin{equation}  \label{eq:angular}
\ell = \frac{1}{i} \int{\Psi^* \frac{\partial\Psi}{\partial\phi}\,dV}.
\end{equation}
The star denotes complex conjugation,
and the frequency $\omega$ is measured in units of $\omega_\perp$.

Extremizing numerically the energy functional (\ref{eq:E-omegaL})
is often not practically convenient because
some of the interesting solutions may not correspond to its minima.
Therefore, we define the Lyapunov functional
\begin{equation}  \label{eq:extended}
  \E2rot = \energy + \frac{a_1}{2}\, (\ell-b_1)^2,
\end{equation}
where we have introduced the constants $a_1, b_1$.
An extremum of $\E2rot$ corresponds to an extremum of $\Erot$
with $\omega= a_1 (b_1-\ell)$.
We find the minima of $\E2rot$ by using a variant of
a numerical norm-preserving
relaxation algorithm \cite{castin} which, in effect, capitalizes on a virial
relation \cite{virial} in order to converge to a wave function of unit norm.
The method is applied on a three-dimensional grid in Cartesian coordinates.
The advantage of using $\E2rot$ is that
we can find stationary points which are not necessarily minima of $\Erot$.
The constants $a_1, b_1$ should be chosen
so as to ensure convergence to the desired
value of the angular momentum, say $\ell_0$.
This usually amounts to choosing $b_1$ close to $\ell_0$,
while $a_1$ should be positive and large enough
so that the stationary point at $\ell_0$ be a minimum.
Once the solutions for a range of angular momenta are known
the frequency is given by $\omega = dE/d\ell$.

\begin{figure}
   \psfig{file=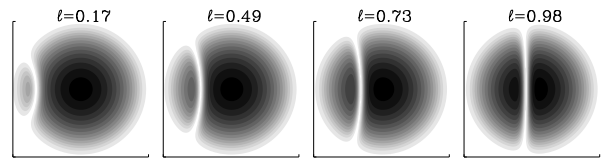,width=8cm}
   \caption{The particle density $|\Psi|^2$ for vortex solutions of the
Gross-Pitaevskii model in a spherical trap ($\beta=1$) at $\delta=85$.
The parameters correspond to the conditions
of the experiment \cite{anderson}.
We show the plane containing the vortex line and the $z$ axis along which
the angular momentum $\ell$ is measured. The images correspond to
four representative values of
$\ell=0.17, 0.49, 0.73, 0.98$ with precession frequencies
$\omega=0.268, 0.221, 0.205, 0.196$ respectively, in units of $\omega_\perp$.
The size of the frames shown is $12 \! \times\! 12$ in units of $a_\perp$.
White colour corresponds to no particles and black colour
to maximum particle density.
}
  \label{fig:cornell_images}
\end{figure}

We first apply the method for parameters similar to those
in Ref.~\cite{anderson}: $\beta=1$ and $\delta=85$.
Since the trap is spherical any arbitrary axis in space
can be chosen as the z-axis along which $\ell$ is measured.
We find vortex states for the range of angular momenta $0<\ell \leq 1$
and we present images of the particle density for
some representative ones in Fig.~\ref{fig:cornell_images}.
The condensate is at the nonrotating spherically symmetric
ground state for $\ell=0$. For $\ell>0$ a vortex line
enters the condensate. As $\ell$ is increased the vortex line
is closer to the center and it is curved for $\ell < 1$;
it becomes the axially symmetric vortex
at the center of the condensate for $\ell=1$.
Following \cite{rosenbusch,aftalion}
we call the vortex mode in Fig.~\ref{fig:cornell_images} the {\it U-vortex}.

\begin{figure}
  \epsfig{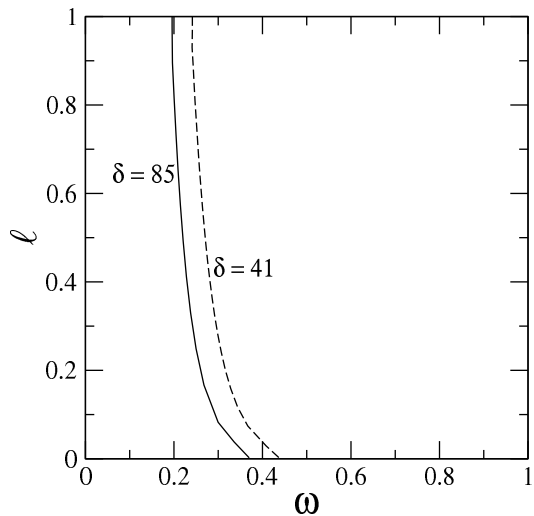}
  \caption{Angular momentum as a function of the precession frequency
for $\beta=1,\, \delta=85$ (solid line)
which corresponds to Fig.~\ref{fig:cornell_images}
(and to experiment \cite{anderson}).
We have a U-vortex branch with precession frequencies $0.20 < \omega < 0.35$
in units of $\omega_\perp$.
We include the corresponding line (dashed)
for a smaller interaction strength $\delta=41$ for comparison.
}
\label{fig:cornell}
\end{figure}

The angular momentum $\ell$ as a function of the
precession frequency $\omega$ of the vortex around the $z$-axis
is shown in Fig.~\ref{fig:cornell}.
We find that single vortex states with $0 < \ell < 1$
have frequencies in the range $0.19 < \omega < 0.35$.
However, except for very low values of $\ell$ the branch has frequencies
in a narrow frequency range, say,
$\omega \approx$ 0.19 to 0.25. In physical units the frequency $f$
defined from $\omega=2\pi f$ takes values in the range
$1.56\, {\rm Hz} < f <  1.95\, {\rm Hz}$
(e.g., $f=1.72$ Hz for $\ell=0.5$);
which is in agreement with the value 1.8 Hz measured
in \cite{anderson}.
In Ref.~\cite{feder} the precession frequency of the vortex is
explained in terms of an anomalous mode of the Bogoliubov spectrum
and it is related to its stability.
The frequencies 1.58 Hz given in the above reference should
be compared with the limiting frequency (1.56 Hz) obtained 
in our calculation.
A somewhat counterintuitive feature of the results presented in Fig.~\ref{fig:cornell}
is that the precession frequency decreases as the angular momentum
increases, which means that vortices closer to the axis precess slower.
This point is discussed further at the end of this section
in connection with stability.

Further in this section
we expand our calculations to axially symmetric elongated traps.
This is an important direction since the majority
of experiments involving vortices have been performed in elongated
condensates.
Previous theoretical work on few-vortex solutions
has focused on very dense elongated condensates
which are practically in the Thomas-Fermi regime
\cite{garcia-ripoll,modugno,aftalion} and satisfy
the condition $\delta \gg 1,\beta$.
In this section we shall focus on the regime $\delta \sim 1$.

\begin{figure}
  \epsfig{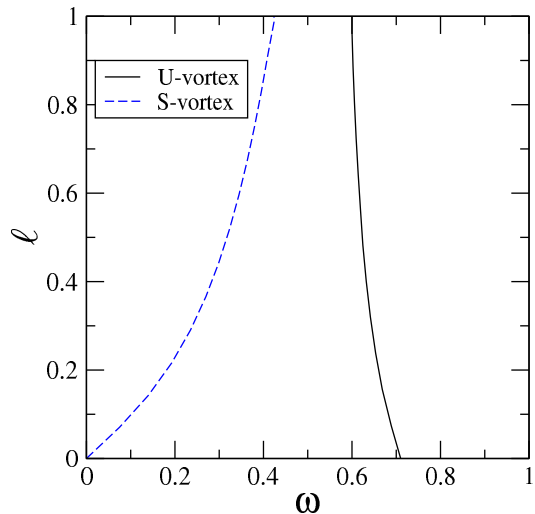}
  \caption{Angular momentum as a function of the precession frequency
for vortices in an axially symmetric elongated trap
with $\beta=1/2,\, \delta=2$.
The solid line shows the U-vortex branch.
The dashed line is the S-vortex branch which contains the solitonic vortex
for $\ell=0$ (at $\omega=0$).
Both modes give the axially symmetric vortex for $\ell = 1$
but this is obtained at different limiting frequencies:
$\omega=0.64$ for the U-vortex and $\omega=0.44$ for the S-vortex.
}
\label{fig:d2}
\end{figure}

\begin{figure}
   \psfig{file=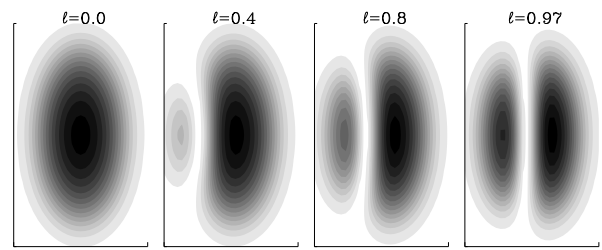,width=8cm}
   \caption{The particle density $|\Psi|^2$ for the U-vortex
corresponding to the solid line in Fig.~\ref{fig:d2} ($\beta=1/2,\, \delta=2$)
for four values of the angular momentum $\ell$.
We show here the plane containing the symmetry axis and the vortex line.
At $\ell=0$ we have the nonrotating ground state and for $\ell=1$
we obtain the axially symmetric vortex.
For $0 < \ell < 1$ the vortex lines are slightly curved.
The size of the frames shown is $6 \! \times\! 10 $ (in units of $a_\perp$).
}
  \label{fig:Uvortex_d2}
\end{figure}

We present here a calculation for $\beta=1/2$ and $\delta=2$.
Using Na atoms and trapping frequencies
$\omega_\perp = 2 \pi\times 7.8$ Hz, $\omega_\| = \omega_\perp/2$,
these values are obtained for $N=27000$ atoms ($N=6700$ for Rb).
In Fig.~\ref{fig:d2} we show by a solid line the angular momentum
for the U-vortex as a function of the precession frequency.
The angular momentum is a decreasing function of $\omega$,
and the precession frequency range over which the U-vortex exists
is narrow and it is located at much higher frequencies compared to
Fig.~\ref{fig:cornell}.
In Fig.~\ref{fig:Uvortex_d2} we present images of the U-vortices
for various values of the angular momentum.
Vortices are off-center and slightly curved.
The axially symmetric vortex is obtained as the limit
for a frequency $\omega_U=0.64$.

\begin{figure}
   \psfig{file=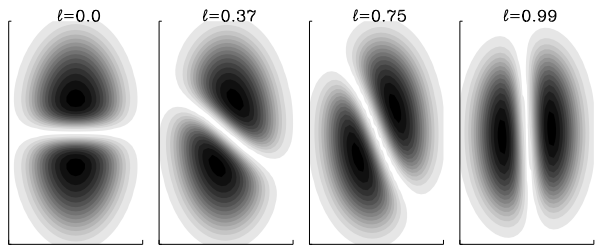,width=8cm}
   \caption{
The particle density $|\Psi|^2$ for the S-vortex
corresponding to the dashed line in Fig.~\ref{fig:d2} ($\beta=1/2,\, \delta=2$)
for four values of the angular momentum $\ell$.
We show here the plane containing the symmetry axis and the vortex line.
At $\ell=0$ we have the {\it solitonic vortex} and for $\ell=1$
we obtain the axially symmetric vortex.
For $0 < \ell < 1$ the vortex lines are slightly curved.
The size of the frames shown is $6 \! \times\! 10 $ (in units of $a_\perp$).
}
  \label{fig:Svortex_d2}
\end{figure}

Refs.~\cite{brand,komineas3,komineas4} study a solution of the
Gross-Pitaevskii equation which has been
called the {\it solitonic vortex}.
This appears to be different than all the vortices discussed
in this section so far.
The solitonic vortex
is an excited state in an elongated non-rotating condensate
which pierces through the trap in a direction perpendicular to
the symmetry axis and therefore
it has angular momentum $\ell=0$ (along the symmetry axis).
An example is shown in the first entry of Fig.~\ref{fig:Svortex_d2}.
The solitonic vortex exists for sufficiently high interaction strengths
and it bifurcates from the dark soliton solution.

We notice that the solitonic vortex breaks spontaneously the
axial symmetry and if
we assume a finite precession frequency $\omega$, it
may acquire a nonvanishing angular momentum $\ell=\ell(\omega)$.
In order to explore this possibility we use the solitonic vortex solution
as an initial guess in our numerical algorithm and
choose appropriate values for the constants in Eq.~(\ref{eq:E-omegaL}).
Representative results for vortex solutions with $0 \leq \ell \leq 1$
are shown in Fig.~\ref{fig:Svortex_d2},
for an elongated condensate ($\beta=1/2$) and for an
interaction strength constant $\delta=2$.
The first entry, for $\ell=0$, is
the solitonic vortex and the next entries show almost straight vortex lines
which are tilted with respect to the $z$ axis.
As the frequency increases the vortex line progressively aligns
with the symmetry axis while its angular momentum increases.
The dependence of $\ell$ on $\omega$
is shown by a dashed line in Fig.~\ref{fig:d2}.
This mode, which may be called the {\it S-vortex} following \cite{aftalion},
gives the axially symmetric vortex ($\ell=1$)
at a limiting frequency $\omega_S = 0.44$
(this value should be compared with Eq.~(\ref{eq:omegaS}) in Sec. III.)
Thus the S and the U-vortex are two different types
of single vortex states in elongated condensates.

The picture summarized in Figs.~\ref{fig:d2}, \ref{fig:Uvortex_d2},
and \ref{fig:Svortex_d2}
is apparently robust for large interaction strengths. In particular,
in the experiments in \cite{rosenbusch} and in
Refs.~\cite{garcia-ripoll,modugno,aftalion}
S and U-vortices are found in condensates which are practically in the
Thomas-Fermi regime ($\delta \gg 1,\beta$).
Unlike in the present case, all vortices were found to be distinctly curved.
Also, in \cite{aftalion} the limiting ($\ell=1$) frequencies
for the axially symmetric vortex appear to coincide
for both the S and the U-vortex.
These differences should be attributed to the
Thomas-Fermi nature of the condensates studied in the above references.

\begin{figure}
  \epsfig{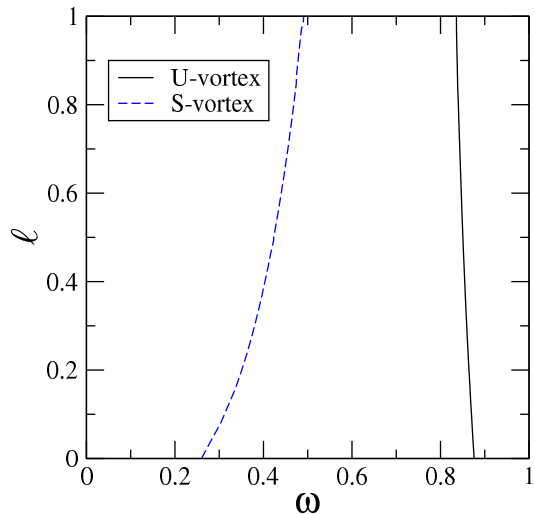}
  \caption{Angular momentum as a function of the precession frequency
for vortices in an axially symmetric elongated trap
with $\beta=1/2,\, \delta=0.28$.
The solid line shows the U-vortex branch.
The dashed line is the S-vortex for which
the $\ell=0$ state (for $\omega\leq 0.26$) is a dark soliton.
Both modes become the axially symmetric vortex at the limit $\ell \to 1$
but this is obtained at different frequencies.
}
\label{fig:a2}
\end{figure}

In order to complete the study of S-vortices
we note that the qualitative picture should be modified for low interaction strengths.
This is because the solitonic vortex is a solution of the nonlinear theory
which exists only above a certain value of the interaction strength,
$\delta > \delta_0$, which depends on $\beta$
\cite{komineas3,komineas4}.
In order to explore the regime $\delta < \delta_0$ we choose
$\delta=0.28$ in the elongated trap with $\beta=1/2$.
Fig.~\ref{fig:a2} shows the angular momentum of the vortex modes
as a function of the precession frequency.
A substantial modification for the S-vortex is
that now $\ell \to 0$ for a finite limiting frequency $\omega = 0.26$.
For $\omega \leq 0.26$ we obtain an axially symmetric wave function with
a node on the $z=0$ plane which, in the context of the present nonlinear theory,
is called a {\it dark soliton} by comparison to the dark soliton of the
nonlinear Schr\"odinger equation.
Summarizing,
the picture in Fig.~\ref{fig:a2} could also be described as following:
a dark soliton which may initially be created in a trap would not respond to
rotations of the system until we exceed a limiting frequency 0.26.
For $\omega > 0.26$ it could turn into an S-vortex which would
become an axially symmetric one for $\omega > 0.49$
(these frequencies should be compared with
Eq.~(\ref{eq:omega}) in Sec. III when setting $\ell=0$
and $\ell=1$ respectively.)
Finally, we note that the particle density plots for the S-vortices
are qualitatively similar to those for higher interaction strengths
presented above (as in Fig.~\ref{fig:Svortex_d2}),
except, of course, for the $\ell=0$ case.

The U-vortex branch is qualitatively similar to its counterpart
for $\delta=2$ but its frequency range is shifted
closer to the transverse trapping frequency ($\omega_\perp$)
while it becomes narrower as $\delta$ decreases.

In Ref.~\cite{virial} a series of virial relations are derived
which should be satisfied by any vortex solution. We have checked
that the solutions presented here have, in particular, the
counterintuitive property that the moments of the particle density
vanish:
$\int{x\,|\Psi|^2\,dV} = \int{y\,|\Psi|^2\,dV} = \int{z\,|\Psi|^2\,dV} = 0$.
This is trivially satisfied by the S-vortex due
to its symmetry along the $z$ axis, but it
is a nontrivial result for the U-vortex.

We comment briefly on the relation of the vortex modes presented
here to vortices in helium II.
In Ref.~\cite{carlson}
an equation is derived for the dynamics of a vortex line
in a rotating bucket containing the superfluid
and it is found that there are vortex lines precessing
on an unstable orbit around
the bucket center at some distance $r$
which is an increasing function of the precession frequency.
Since the angular momentum of the vortex is a decreasing function
of $r$ (as discussed in \cite{bauer}), we conclude that the angular momentum
is a decreasing function of the precession frequency,
that is the situation is analogous to that for the U-vortex
in Figs.~\ref{fig:cornell},\ref{fig:d2}, and \ref{fig:a2}.

We finally turn to the delicate issue of stability
of the calculated vortex states.
According to \cite{butts} ``mechanical stability'' requires 
that the energy be a convex function of angular momentum:
$d\omega/d\ell = d^2 E/d\ell^2 > 0$.
However, this is a condition for stability of the solution in the
rotating frame under variations that
change the angular momentum. In experiments in which the trap is
axially symmetric (\ie, when the system is no longer being
stirred), the angular momentum is conserved and this condition is
irrelevant for the stability of the modes: what is relevant is that
the mode should be {\it dynamically stable} against decay into other
modes which have the same angular momentum.  As is
evident from Fig.~\ref{fig:cornell} the precessing U-vortex has
$d\omega/d\ell < 0$, so it does not satisfy the criterion of mechanical
stability of Ref.~\cite{butts}. However,
the U-vortex must be dynamically stable since it is the lowest
energy configuration for given angular momentum $0<
\ell < 1$ as we know from the calculations in the present section.
Indeed the U-vortex has been seen in
experiments \cite{anderson} in which angular momentum is conserved.

It is also interesting to comment on the issue of overall
{\it thermodynamic} stability of the rotating system.
A single axially symmetric ($\ell=1$) vortex will have lower energy
in the rotating frame than the nonrotating ground state if
$\omega < E_V - E_0$ where $E_V$ is the vortex
energy and $E_0$ is the ground state energy \cite{donnelly}.
Now, taking into account that $\omega=dE/d\ell$, we may write
\begin{equation}  \label{eq:ev}
 \energy_V - \energy_0 = \int_0^1{\omega(\ell)\,d\ell}.
\end{equation}
Applying this relation for the U-vortex calculated for a spherical
trap with $\delta=85$ (see Fig.~\ref{fig:cornell}) we find
$\energy_V - \energy_0 = 0.23$ or, in physical units,
$1.8$ Hz which lies in the range of precession frequencies
for the U-vortex.

\section{Dilute condensates}

\begin{figure}
  \epsfig{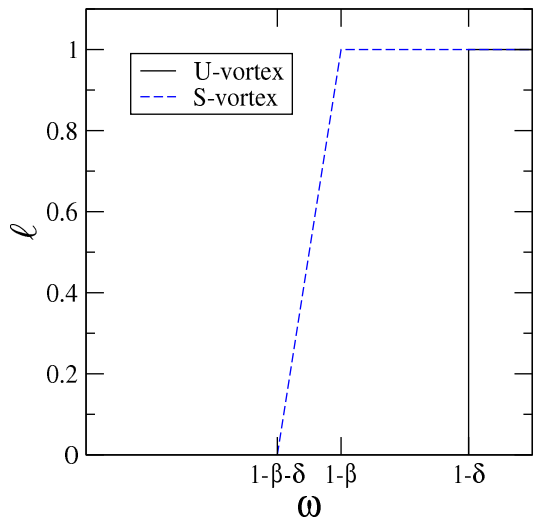}
  \caption{ 
The angular momentum $\ell$ as a function of the frequency
for $\delta \ll 1,\beta,1-\beta$ for single vortex states. The solid
line is the result of Ref.~\cite{butts}. The dashed line
is a plot of Eq.~(\ref{eq:omega}).
The horizontal axis is not to scale.
The wave function at $\ell=0$
for the dashed line ($\omega \leq 1-\beta-\delta$) is $\Psi_1$,
while for the solid line ($\omega \leq 1-\delta$) it is the Gaussian.
The $\ell=1$ state is $\Psi_2$ for both cases.
}
\label{fig:LLL}
\end{figure}

We now turn to discuss more specifically dilute condensates which
offer the possibility to study vortex modes
in the limit of a small interaction strength where the energy levels
of the system are close to those of the harmonic oscillator.
This analysis will provide an analytic description of the features
of the S and U-vortices.
In Ref.~\cite{butts} the energy $\Erot$ of a rotating system is minimised
in the case of the very dilute limit in the sense that the nonlinear
term in Eq.~(\ref{eq:energy}) is much smaller than the other
energy scales, namely $\hbar\omega_\perp,\hbar\omega_\|$, or
\begin{equation}  \label{eq:constraint}
 \delta \ll 1,\beta.
\end{equation}
The problem is then reduced to the 
lowest Landau level (LLL) on the transverse plane ($\delta \ll 1$)
while the wave function in the longitudinal direction $z$
is the Gaussian ground state of the 1D harmonic oscillator.
We refer to this as the 2D LLL regime.
An analytic solution for vortices with angular momentum in the
range $0 < \ell \le 1$ is given in \cite{vorov}.
The minimisation of $\Erot$ gives
the Gaussian nonrotating ground state as the global minimum for
$\omega < \omega_U \equiv 1-\delta$.
For $\omega > \omega_U$ the axially symmetric vortex
becomes the global minimum of the functional (\ref{eq:E-omegaL}).
Exactly at $\omega=\omega_U$ there is a set of states with $0 < \ell < 1$
which are off-center vortices.
These are denoted by a solid line in Fig.~\ref{fig:LLL}
which shows the angular momentum as a function of the
precession frequency.
For higher values of $\omega$ (but still lower than the
transverse frequency of the potential) many-vortex
states and vortex lattices become the absolute minimum \cite{butts,vorov}.
All vortex lines are
parallel to the symmetry axis of the trap since this is a 2D calculation.
For larger values of $\delta$ these evolve in the curved vortex lines as,
\eg, in Fig.~\ref{fig:cornell_images} and thus they correspond to
the U-vortex mode discussed in the previous section. 

In elongated traps excited states of the rotating
condensate can be found without mixing higher Landau levels.
This is an interesting issue also because most experiments involving vortices,
and in particular those in which the shapes of individual vortex lines
were studied, were performed in very elongated traps,
that is at $\beta \ll 1$ \cite{rosenbusch}.
We first consider
the case of a non-interacting gas, \ie, $\delta=0$,
whose first excited state has a node in the $z$ direction:
\begin{equation}  \label{eq:soliton}
  \Psi_1 = \frac{\sqrt{2}}{\pi^{3/4}}\, z \, e^{-(\rho^2+ z^2)/2},
\end{equation}
its energy is $\energy_1=\beta$, and has vanishing angular momentum.
We note that for increasing $\delta$ the wave function $\Psi_1$ evolves
to the dark soliton in the framework of the nonlinear Gross-Pitaevskii model.
A further excited state is of the vortex type:
\begin{equation}  \label{eq:vortex}
\Psi_2 = \frac{1}{\pi^{3/4}}\, \rho\, e^{i\phi} \, e^{-(\rho^2+ z^2)/2},
\end{equation}
and it has energy $\energy_2=1$ and angular momentum $\ell=1$.
It is thus obvious that for the frequency
\begin{equation}  \label{eq:omegaS}
 \omega_S = 1 - \beta
\end{equation}
the two states (\ref{eq:soliton}) and (\ref{eq:vortex}) have the same
energy $\Erot$ in the rotating frame.
For $\omega < \omega_S$ we have $\Psi_1$ as the first excited state
while $\Psi_2$ becomes the first excited state for $\omega > \omega_S$.

The simple picture is modified in the
presence of a weak interaction term, that is, for $\delta$ small but finite.
For $\delta \ll \beta, 1-\beta$ we make the ansatz
\begin{equation}  \label{eq:twostates}
 \Psi = c_1\,\Psi_1 +  c_2\,\Psi_2,
\end{equation}
where the coefficients $c_1, c_2$  are constrained by $|c_1|^2 + |c_2|^2= 1$
and $|c_2|^2=\ell$.
Substituting $\Psi$ in (\ref{eq:energy}) we find the energy
of the first excited state as a function of the angular momentum $\ell$:
\begin{equation}  \label{eq:energy2}
\energy(\ell) =
\frac{\delta}{2} \ell^2 + (1-\beta-\delta) \ell + \beta + \frac{3\delta}{2},
\end{equation}
where we have subtracted the ground state energy of the harmonic oscillator
$1+\beta/2$, a convention that will be adopted from now on.
Now, $\Erot(\ell)$ has a minimum for any $\ell$ in the range
$0 < \ell < 1$ and a frequency $\omega=dE/d\ell$ or
\begin{equation}  \label{eq:omega}
 \omega = 1-\beta - \delta (1-\ell),
\end{equation}
as depicted by the dashed line in Fig.~\ref{fig:LLL}.
For $\omega \leq  1-\beta-\delta$
the wave function is $\Psi=\Psi_1$ ($\ell=0$) while
for $\omega \ge  1-\beta$ we have $\Psi=\Psi_2$ ($\ell=1$).

The vortex line represented by the ansatz (\ref{eq:twostates})
is found by setting
\begin{eqnarray}  \label{eq:tilted}
\Psi=0 & \Rightarrow & \sqrt{2}\, c_1\, z +  c_2 (x + i y) = 0 \nonumber \\
& \Rightarrow & y=0,\; \sqrt{2}\,c_1\, z +  c_2\, x = 0,
\end{eqnarray}
where for simplicity we may consider $c_1, c_2$ real.
For frequencies in the range $ 1-\beta < \omega < 1-\beta-\delta$
we have $c_1, c_2 \neq 0$ and thus
Eq.~(\ref{eq:tilted}) gives a straight vortex line
which is tilted with respect to the symmetry axis $z$.
Summarizing, the first excited state (\ref{eq:twostates}) has a
wave function with a nodal plane perpendicular to the $z$ axis
for low frequencies which turns to a tilted vortex for 
$ 1-\beta-\delta < \omega < 1-\beta$.
This gradually aligns
with the symmetry axis as the frequency increases.
It becomes the axially symmetric vortex above the limiting 
frequency $\omega \ge \omega_S = 1-\beta$.
This is the weak interaction limit of the
S-vortex discussed in the previous section.

Both the S and the U-vortex become the axially symmetric
vortex for a sufficiently high $\omega$. However, the corresponding
limiting frequencies for the two types of vortices do not coincide.
In fact, there is no sense of precession in an axially symmetric vortex
and the limiting frequency may be absorbed into an effective
chemical potential $\bar{\mu} = \mu + \omega$
while the wave function reduces to a quasi static configuration
with chemical potential $\bar{\mu}$.

We are able to take our calculation one step further
if we assume a very elongated trap, \ie, $\beta \ll 1$,
and an interaction strength comparable to the latter energy scale:
\begin{equation}  \label{eq:3DLLL}
 \delta \ll 1,\quad {\rm and}\quad \delta \sim \beta.
\end{equation}
In this case it is useful to extend the LLL calculation to 3D
by including higher axial oscillator states.
We use the wave function ansatz
\begin{equation}  \label{eq:ansatz}
\Psi(\rho,z,\phi) = \sum c_{mn} \chi_m(\rho,\phi)\, \xi_n(z),
\end{equation}
where $c_{mn}$ are complex constants,
\begin{equation}  \label{eq:chi}
\chi_m(\rho,\phi) = \left( \frac{1}{\pi m!}\right)^{1/2}\,
e^{i m \phi}\, \rho^m\, e^{-\rho^2/2}
\end{equation}
are eigenstates of the 2D harmonic oscillator and
\begin{equation}  \label{eq:xi}
 \xi_n(z) = \left( \frac{2^n}{(2 n-1)!!}\right)^{1/2}\,\frac{1}{\pi^{1/4}}\;
  z^n\, e^{-z^2/2}
\end{equation}
form a basis for the expansion of the wave function in the axial direction.
We aim to find the vortex states for a fixed angular momentum within
the present ansatz.
An efficient way to do this is by minimizing
the Lyapunov functional
\begin{equation}  \label{eq:extended2}
  \ErotMM = \energy 
 + \frac{a_1}{2}\, (\ell-b_1)^2 + \frac{a_2}{2}\, (\nu-b_2)^2,
\end{equation}
where $\ell$ is the angular momentum functional of Eq.~(\ref{eq:angular}) and
$\nu \equiv \int |\Psi|^2\,dV$.
The constants $a_1, a_2, b_1, b_2$
are chosen so as to obtain a solution for the desired
number of particles and angular momentum.
We substitute the wave function (\ref{eq:ansatz}) in (\ref{eq:extended2})
and minimise $\ErotMM$ with respect to $c_{mn}$
by a conjugate gradient method.
Computer limitations allow for modes typically up to $m=4$ and $n=14$.

\begin{figure}
  \epsfig{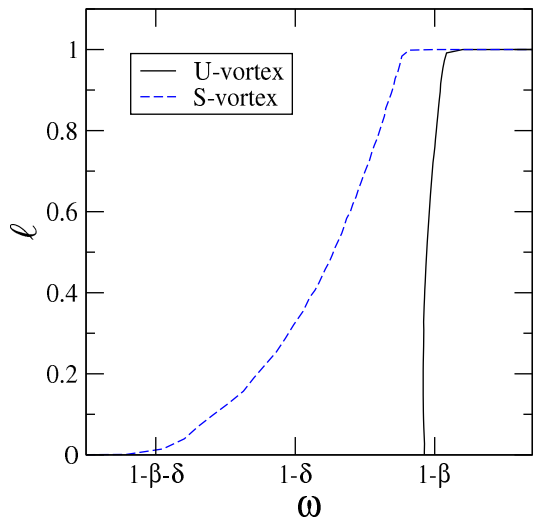}
  \caption{
The angular momentum $\ell$ as a function of the frequency
for the S (dashed line) and U-vortex (solid line)
within the 3D LLL calculation for $\delta/\beta=2$.
Here, $d\ell/d\omega >0$ (except for small values of $\ell$)
unlike in Fig.~\ref{fig:LLL}.
}
\label{fig:3DLLL}
\end{figure}

When the interaction term
is small, \ie, $\delta \ll \beta$ the algorithm converges to
the two single-vortex solutions which
were discussed in Fig.~\ref{fig:LLL}. Specifically, for the U-vortex we have
that all $c_{mn}$ with $n\neq 0$ vanish.
For the S-vortex we have that only $c_{01}, c_{10}$ are nonzero.
When $\delta$ becomes comparable or greater than $\beta$
several $c_{mn}$ are nonzero
and their values decrease with respect to both indices.
For the S-vortex we find that all $c_{mn}$ with $m+n$ even vanish,
and this is consistent with the symmetry of an S-shaped vortex.
For the U-vortex we find that for all $n$ odd $c_{mn}=0$
which is consistent with the symmetry of a U-shaped vortex.

Fig.~\ref{fig:3DLLL} shows the angular momentum of the two vortices
as a function of precession frequency
for parameter values $\delta/\beta=2$.
The most dramatic effect is seen in the U-vortex
where $d\ell/d\omega$ is now positive except for
small values of $\ell$.
This means that the abrupt transition from the
$\ell=0$ to the $\ell=1$ state depicted in Fig.~\ref{fig:LLL}
is smoothed out and the angular momentum of the U-vortex is now a smooth
function of the frequency $\omega$, although
the range of frequencies corresponding to $0<\ell<1$ is narrow.
The vortices in the latter range are located off-center while
the contribution of modes with $n>0$ in the wave function (\ref{eq:ansatz})
means that vortex lines are curved.
The  S-vortex lines are also curved
and the slope of the branch (dashed line in Fig.~\ref{fig:3DLLL})
is not constant unlike in Fig.~\ref{fig:LLL}.

\begin{figure}
  \epsfig{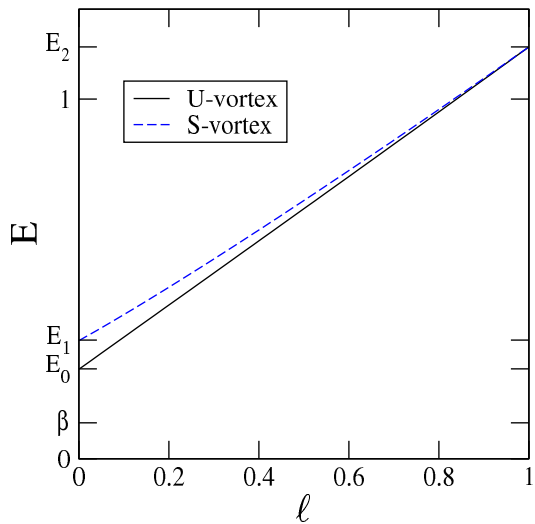}
  \caption{
The energy $\energy$ of Eq.~(\ref{eq:energy})
as a function of the angular momentum $\ell$
for the S (dashed line) and the U-vortex (solid line) of Fig.~\ref{fig:3DLLL}
where $\delta/\beta=2$.
The energies of the nonrotating ground state, and of
the wave functions $\Psi_1$,
and $\Psi_2$, are $\energy_0,\,\energy_1$, and $\energy_2$ correspondingly.
(In the noninteracting system the corresponding energies are $0,\,\beta$, and 1,
having subtracted the ground state energy of the harmonic oscillator
$1+\beta/2$.)
}
\label{fig:EvL_3DLLL}
\end{figure}

A concise presentation of the main features of the two vortex modes
can be given by the graph of the energy as a function of the
angular momentum shown in Fig.~\ref{fig:EvL_3DLLL}.
The energy of the U-vortex mode for $\ell=0$ is the energy of the
nonrotating ground state. The corresponding value for the S-vortex
is the energy of the wave function $\Psi_1$.
The S-vortex has higher energy than the U-vortex but both curves
converge to the same value for $\ell=1$ which is the energy of the axially
symmetric vortex $\Psi_2$.

\section{Conclusions}

We have analysed the two single vortex modes, the S-vortex
and the U-vortex, which are stationary states of the Gross-Pitaevskii
model precessing at a constant frequency $\omega$
in a BEC confined in an axially symmetric harmonic potential.
Vortices for the whole range of
angular momenta $0 \leq \ell \leq 1$ have been studied.
The precession frequency range is distinctly different
for the two modes. The U-vortices fall within a relatively
narrow frequency range. For those U-vortices studied in Sec. II
we find that the precession frequency is a decreasing
function of the angular momentum.
The S-vortex is an excited state in elongated condensates.
Their frequency range extends down to zero
(for sufficiently high interaction strengths)
and the nonrotating member of the family is the so-called
solitonic vortex which is a solution of the nonlinear
Gross-Pitaevskii model \cite{komineas3,brand}.
Both the U-vortex and the S-vortex have been directly seen in experiment
\cite{anderson,rosenbusch}.
Perhaps the S-vortex is also related to
shape deformations of vortex lines observed in the form
on Kelvin modes \cite{bretin,smith}.

A complete analysis for vortex solutions
could be carried out for a dilute condensate
within a 3D lowest Landau level calculation which includes
the axial oscillator states.
We have given a derivation for the S-vortex
which is complementary to the LLL calculation
for the U-vortex \cite{butts}.
Including the axial oscillator states causes the transition from
a nonrotating ground state to an axially symmetric vortex to happen smoothly
as the rotation frequency increases:
vortices with angular momentum $\ell$ less than unity are
thermodynamically stable.

Both the S and the U-vortex become axially symmetric
in the limit $\ell \to 1$ but this happens at different frequencies
in the two cases.
This seems to be in disagreement with the results reported
in \cite{aftalion}, however, one should note that in the latter reference
very elongated and very dense condensates are studied.
Although the results are related, an extrapolation of our results to the
Thomas-Fermi regime, and thus a direct comparison to \cite{aftalion},
does not appear straightforward.
We further find that, for weak interactions,
an S-vortex line is only slightly bent
in contrast to the original observations in \cite{rosenbusch}.
In fact,
both vortex modes in the dilute limit presented in
Fig.~\ref{fig:LLL} represent straight vortex lines.
However, as the density increases vortex lines tend to bend
(as, \eg, in Figs.~\ref{fig:Uvortex_d2},\ref{fig:Svortex_d2}).
Thus, the difference between the present results and the observations
is clearly due to the difference in the density of the condensates.

\begin{acknowledgments}
S.K. is grateful to the ENS group for hospitality and
for discussions.
S.K. and N.R.C are grateful to the Kavli Institute of Theoretical Physics
in Santa Barbara for hospitality and acknowledge discussions
during the ``Quantum gases'' program from which this work has benefited.
This work was supported by
EPSRC Grant Nos GR/R96026/01 (SK) and GR/S61263/01 (NRC).
\end{acknowledgments}

\end{document}